\pdfoutput=1
\documentclass[longauth]{aa} 

\usepackage{lineno}
\usepackage{multicol}

\usepackage{natbib,twoopt}
 \usepackage[breaklinks=true]{hyperref} 
\bibpunct{(}{)}{;}{a}{}{,}
  \makeatletter
   \newcommandtwoopt{\citeads}[3][][]{\href{http://adsabs.harvard.edu/abs/#3}%
     {\def\hyper@linkstart##1##2{}%
      \let\hyper@linkend\@empty\citealp[#1][#2]{#3}}}
   \newcommandtwoopt{\citepads}[3][][]{\href{http://adsabs.harvard.edu/abs/#3}%
     {\def\hyper@linkstart##1##2{}%
      \let\hyper@linkend\@empty\citep[#1][#2]{#3}}}
   \newcommandtwoopt{\citetads}[3][][]{\href{http://adsabs.harvard.edu/abs/#3}%
     {\def\hyper@linkstart##1##2{}%
      \let\hyper@linkend\@empty\citet[#1][#2]{#3}}}
   \newcommandtwoopt{\citeyearads}[3][][]%
     {\href{http://adsabs.harvard.edu/abs/#3}
     {\def\hyper@linkstart##1##2{}%
      \let\hyper@linkend\@empty\citeyear[#1][#2]{#3}}}
 \makeatother

\newcommand{\e}{\epsilon}
\newcommand{\g}{\gamma}
\newcommand{\gp}{\gamma^{\prime}}

\newcommand{\psim}{\lower.5ex\hbox{$\; \buildrel \propto \over\sim \;$}}
\newcommand{\lbar}{\lower.0ex\hbox{$\; \buildrel
{\lower0.0ex \hbox{-}} \over\lambda  \;$}}

\newcommand{\cm}{\mathrm{cm}}

\newcommand{\erg}{\mathrm{erg}}

\newcommand{\s}{\mathrm{s}}

\newcommand{\fermilat}{\emph{Fermi}~LAT~}
\newcommand{\pks}{PKS~1424$-$418}

\usepackage{graphicx}
\usepackage[varg]{txfonts}
  \usepackage{color}  

\begin{document}
   \title{Unusual Flaring Activity in the Blazar \pks\ During 2008-2011}
   \titlerunning{Flaring Activity in \pks\ During 2008-2011}
\authorrunning{Buson et al.}
\author{
S.~Buson$^{(1,2)}$ \and 
F.~Longo$^{(3,4)}$ \and 
S.~Larsson$^{(5,6,7)}$ \and 
S.~Cutini$^{(8,9)}$ \and
 J.~Finke$^{(10)}$ \and 
S.~Ciprini$^{(8,9)}$ \and 
R.~Ojha$^{(11)}$ \and
F.~D'Ammando$^{(12)}$ \and 
D.~Donato$^{(13,14)}$ \and
D.~J.~Thompson$^{(11)}$ \and 
R.~Desiante$^{(3)}$ \and
D~Bastieri$^{(1,2)}$ \and
S.Wagner$^{(15)}$ \and M.Hauser$^{(15)}$ \and
L.~Fuhrmann$^{(16)}$ \and
M.~Dutka$^{(17)}$ \and
C.~M\"uller$^{(18,19)}$ \and
M.~Kadler$^{(12,18,19,20,21)}$ \and
E.~Angelakis$^{(16)}$ \and J. A.~Zensus$^{(16)}$ \and
J.~Stevens$^{(22)}$\and
J.~M. Blanchard$^{(23)}$\and
P.~G. Edwards$^{(24)}$\and
J.E.J.~Lovell$^{(23)}$ \and
M.A.~Gurwell$^{(25)}$\and
A.E.~Wehrle$^{(26)}$ \and
A.~Zook$^{(27)}$
}

\institute{
\inst{1}~Istituto Nazionale di Fisica Nucleare, Sezione di Padova, I-35131 Padova, Italy  \email{sara.buson@pd.infn.it} \\
\inst{2}~Dipartimento di Fisica e Astronomia ``G. Galilei'', Universit\`a di Padova, I-35131 Padova, Italy\\ 
\inst{3}~Istituto Nazionale di Fisica Nucleare, Sezione di Trieste, I-34127 Trieste, Italy \email{francesco.longo@trieste.infn.it} \\
\inst{4}~Dipartimento di Fisica, Universit\`a di Trieste, I-34127 Trieste, Italy\\ 
\inst{5}~Department of Physics, Stockholm University, AlbaNova, SE-106 91 Stockholm, Sweden\\ 
\inst{6}~The Oskar Klein Centre for Cosmoparticle Physics, AlbaNova, SE-106 91 Stockholm, Sweden\\ 
\inst{7}~Department of Astronomy, Stockholm University, SE-106 91 Stockholm, Sweden\\ 
\inst{8}~Agenzia Spaziale Italiana (ASI) Science Data Center, Via del Politecnico, 00133  (Roma), Italy\\
\inst{9}~Istituto Nazionale di Astrofisica - Osservatorio Astronomico di Roma, I-00040 Monte Porzio Catone (Roma), Italy\\ 
\inst{10}~Space Science Division, Naval Research Laboratory, Washington, DC 20375-5352, USA\\ 
\inst{11}~NASA Goddard Space Flight Center, Greenbelt, MD 20771, USA\\ 
\inst{12}~INAF Istituto di Radioastronomia, I-40129 Bologna, Italy\\
\inst{13}~Center for Research and Exploration in Space Science and Technology (CRESST) and NASA Goddard Space Flight Center, Greenbelt, MD 20771, USA\\
\inst{14}~Department of Physics and Department of Astronomy, University of Maryland, College Park, MD 20742, USA\\ 
\inst{15}~Landessternwarte, Universitat Heidelberg, Konigstuhl, D-69117 Heidelberg, Germany\\
\inst{16}~Max-Planck-Institut f\"ur Radioastronomie, Auf dem H\"ugel 69, 53121 Bonn, Germany\\
\inst{17}~The Catholic University of America, 620 Michigan Ave., N.E.  Washington, DC 20064\\
\inst{18}~Institut f\"ur Theoretische Physik and Astrophysik, Universit\"at W\"urzburg, D-97074 W\"urzburg, Germany\\ 
\inst{19}~Dr. Remeis-Sternwarte Bamberg, Sternwartstrasse 7, D-96049 Bamberg, Germany\\ 
\inst{20}~Erlangen Centre for Astroparticle Physics, D-91058 Erlangen, Germany\\ 
\inst{21}~Universities Space Research Association (USRA), Columbia, MD 21044, USA\\ 
\inst{22} CSIRO Astronomy and Space Science, ATNF, Locked Bag 194, Narrabri NSW 2390, Australia\\
\inst{23} School of Mathematics \& Physics, Private Bag 37, University of Tasmania, Hobart TAS 7001, Australia\\
\inst{24} CSIRO Astronomy and Space Science, ATNF, PO Box 76, Epping NSW 1710, Australia \\
\inst{25} Harvard-Smithsonian Center for Astrophysics, Cambridge, MA, USA \\
\inst{26} Space Science Institute, Boulder, CO, USA \\
\inst{27} Jet Propulsion Laboratory, California Institute of Technology, CA 91109 Pasadena, USA \\
}

\date{Received 2014 January 2 / Accepted 2014 June 28}

  \abstract
{Blazars are a subset of  active galactic nuclei (AGN) with jets that are oriented along our line of sight.
Variability and spectral energy distribution (SED)  studies are crucial tools for understanding the physical processes responsible for  observed AGN emission.}
{We report peculiar behaviour in the bright $\gamma$-ray blazar \object{PKS~1424$-$418} and use its strong variability 
to reveal information about the particle acceleration and interactions in the jet.} 
{Correlation analysis of the extensive optical coverage by the ATOM telescope and nearly continuous $\gamma$-ray coverage by the \emph{Fermi} Large Area Telescope is combined with broadband, time-dependent modeling of the SED incorporating supplemental information from radio and X-ray observations of this blazar.}
{We analyse in detail four bright phases at optical-GeV energies. 
These flares of \pks\ show high correlation between these energy ranges, with the exception of one large optical flare that coincides with relatively low $\gamma$-ray activity.
Although the optical/$\gamma$-ray behaviour of \pks\ shows variety, 
the multiwavelength modeling indicates that these differences can largely be explained by changes 
in the flux and energy spectrum of the electrons in the jet that are radiating.
We find that for all flares the SED is adequately represented  
by a leptonic model that includes inverse Compton emission from 
external radiation fields with similar parameters.
}
{Detailed studies of individual blazars like \pks\ during periods of enhanced activity in different wavebands are helping us identify underlying patterns in the physical parameters in this class of AGN.}

 \maketitle

\section{Introduction}
Blazars are radio-loud active galactic nuclei (AGN) with relativistic
jets aligned close to our line of sight \citep[e.g.,][]{blandford78}.
With very few exceptions \citep[e.g., 4C+55.17;][]{4C55.17}
they exhibit variable emission at all wavelengths, from radio to
$\gamma$ rays, on time scales as short as hours or even minutes
\citep{pks2155_hess_07,pks2155_hess_09,4C21.35_magic}.  
Their spectral energy distributions (SEDs) in the $\nu \mathrm{F}_\nu$ representation 
display two broad bumps  usually attributed to 
synchrotron and inverse Compton processes: 
the first bump is located at infrared-optical frequencies but in some sources it can extend to X-ray frequencies, 
while the second bump is found at X-ray/$\gamma$-ray frequencies. 
The frequency of the lower energy peak ($\nu_{sync}$) has been used to subdivide
blazars into three different classes: high, intermediate, and low synchrotron peaked 
(HSP, ISP and LSP, respectively) depending on whether 
 $\nu_{sync}>10^{15}$\ Hz, $10^{14}\ \mathrm{Hz}<\nu_{sync}<10^{15}$\ Hz or $\nu_{sync}<10^{14}$\ Hz \citep{bsl}.
For LSP blazars,
especially flat spectrum radio quasars (FSRQs; blazars with strong,
broad emission lines) it is common for the optical variations to be
correlated with the $\sim$\ GeV $\gamma$\ rays
\citep[e.g.,][]{bonning09,chatterjee11} indicating that emissions in
these wavebands probably originate from the same electron population.  This correlated variability reflects what one would
expect for a model where the low-energy bump is created by synchrotron
emission and the high-energy one is created by Compton scattering of
some soft seed photon source. This idea is also reflected in their  SEDs, in which the optical emission is generally seen
on the decreasing side of the low-energy bump and the GeV $\gamma$
rays fall on the decreasing side of the high-energy bump.  

Reality, however, can be more complicated; a wide variety of blazar
variability behaviours is observed, some of which
can be difficult to explain in this simple picture.  In some instances
for LSPs, the optical and $\gamma$-ray components show correlated
variability, but the optical has smaller flux variations than the $\g$
rays.  This has been interpreted as contamination in the optical band
by an underlying accretion disk \citep{bonning09}.  Although often the
$\gamma$-ray and optical flares are simultaneous
\citep[e.g.,][]{marscher10}, at other times the optical flare occurs
before the $\gamma$-ray flare \citep{marscher10} or lags behind the 
$\gamma$-ray flare \citep{bonning09,LAT3C279}.

In recent years, FSRQ \pks\  (\object{J1427-4206}) has presented an excellent opportunity to use multiwavelength variability as a probe of blazar behaviour, for several reasons:


\begin{enumerate}
\item The Automatic Telescope for Optical Monitoring \citep[ATOM;][]{ATOM09}
provides dense optical light curve coverage for \pks.
\item This optical coverage complements the survey mode of the Large Area Telescope (LAT) 
on the {\em Fermi Gamma-Ray Space Telescope} \citep{LATinstrument}. 
\item \pks\ exhibited two major flares in 2009-2010, seen in both optical \citep{ATOMAtel09, ATOMATel10} and
$\gamma$ rays \citep{LATATel09, LATAtel10} and largely isolated from
other activity.  The clear pattern displayed during the outbursts optimized correlation studies.
\item With mean flux densities over 3\,Jy and known variability at multiple radio frequencies 
\citep{tingay}, \pks\  is a strong and variable radio source. 
High resolution Very Long Baseline Interferometry (VLBI) observations of this blazar are being carried out
as part of the TANAMI program  \citep{Ojha2010}, and radio monitoring is being done as part of the F-GAMMA project \citep{Fuhrmann2007,Angelakis2008}, the Submillimeter Array \citep[SMA,][]{Gurwell07}, and other radio telescopes.
\item The instruments on the \emph{Swift} satellite \citep{Gehrels2004} have provided additional optical, UV, and X-ray coverage of \pks.   
\end{enumerate}

\pks\ has a redshift of $z=1.522$ \citep{white88}
giving it a luminosity distance\footnote{In a flat
$\Lambda$CDM cosmology where $H_0=67.11$\ km s$^{-1}$ Mpc$^{-1}$,
$\Omega_m=0.3175$, and $\Omega_\Lambda=0.6825$ \citep{plankcosmo}.}
of $d_L=11.3$\ Gpc.  Although $\gamma$ radiation from \pks\ was reported in the
Third EGRET Catalog as 3EG J1429$-$4217 \citep{3EGcatalog}, it was not
included in the \fermilat Bright Source List, based on data from
the first three months of the mission \citep{LATBSL}. 
The blazar was then reported in the 
First \citep[1FGL,][]{LAT1FGL} and Second \citep[2FGL,][]{nolan12} LAT Catalogs as as  \object{1FGL J1428.2$-$4204} and \object{2FGL J1428.0$-$4206}, respectively.

  In this work we report the results of the LAT and ATOM monitoring of \pks\ accompanied 
  by multiwavelength observations across the electromagnetic spectrum. 
  The paper is structured as follows.
  In section \ref{sec:multi} we present the collection of multiwavelength data used in this study.  
  Section \ref{sec:lc} describes the multiwavelength light curve, showing the isolated flares that are the focus of this work, while 
  Section \ref{sec:var} presents a detailed cross-correlation analysis of the optical and $\gamma$-ray data.
  Section \ref{sec:sed} describes the modeling of the broadband SED, 
  and Section \ref{sec:conclu} presents the conclusions.

\section{Multiwavelength data}\label{sec:multi}

\subsection{ATOM optical observations}

\begin{figure*}[!ht]
\centering
\leftskip-0.7cm
 \includegraphics[width=1.1\textwidth]{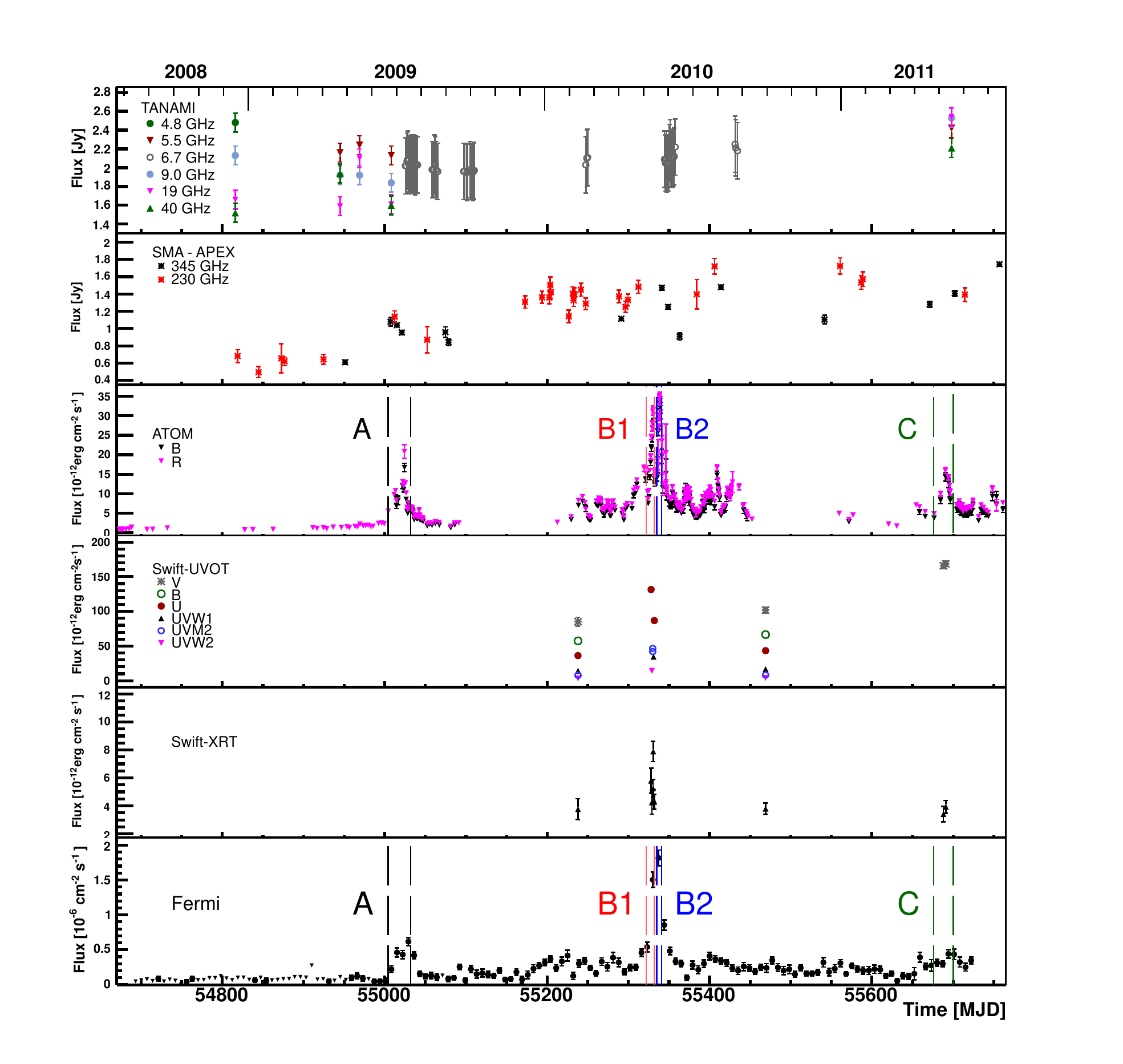}
\vspace{-1.0cm}
 \caption{Multiwavelength light curve of \pks. From top to bottom: 
 first panel displays radio data from TANAMI in several bands between 4.8 and 40\,GHz;
second panel displays 230\,GHz (red symbols) and 345\,GHz (black symbols) data from SMA and APEX, respectively;
third panel displays optical B-band (red points) and R-band (green points) data from ATOM, \textit{Swift}-UVOT data (W1, M2, W2, and V, B, U filters); 
fourth panel displays \textit{Swift}-XRT data;
fifth panel displays \fermilat  flux (E$>$100 MeV).
Vertical lines denote the  prominent outbursts
 studied in this work: black flare A, red flare B1, blue flare B2 and green flare C.}
 \label{mwllc}%
 \end{figure*}

The 75-cm telescope ATOM~\citep{Hauser2004}, located 
in Namibia, monitors the flux of \pks\ in  two different
filters: B (440 nm) and R (640 nm) 
according to \cite{Bessel1990}.  ATOM is operated robotically by the High Energy Stereoscopic System (H.E.S.S.) 
collaboration and obtains automatic observations of confirmed or potentially $\gamma$-ray-bright blazars. A 4$\arcsec$ radius aperture is used for both filter bands.  Data analysis (debiassing, flat fielding, and photometry with Source-Extractor, Bertin \& Arnouts 1996) is conducted automatically using our own pipeline.  The ATOM results are shown in panel 3 of Fig. \ref{mwllc}, which summarizes the multi-wavelength temporal flux behaviour of \pks. 

  \subsection{Fermi LAT gamma-ray observations}

\fermilat is a pair-conversion telescope optimized for energies from 20 MeV  to  greater than 300 GeV  \citep{LATinstrument}. 
Taking advantage of the LAT's large field of view ($\sim$2.4 sr), the \fermilat observatory operated in 
scanning mode provides coverage of the full sky every three hours and offers a good opportunity to follow \pks\ at $\gamma$-ray energies.

We analysed the data sample, which covers observations from 2008 August 4 (MJD 54682)
to 2011 June 13 (MJD 55725),  with the standard \textit{Fermi Science Tools} (version v9r32p5), the P7REP\_SOURCE\_V15 LAT Instrument Response Functions
(IRFs), and associated diffuse emission models\footnote{The P7REP data, IRFs, and diffuse models (gll\_iem\_v05.fit and iso\_source\_v05.txt) are available at http://fermi.gsfc.nasa.gov/ssc/.}.

  The data selection was based on the following criteria.
  Only good quality P7REP source class events \citep{bregeon} 
were extracted from a circular region of interest (ROI) of $10^{\circ}$ radius centered at the location of 
 \pks\ and considered in the analysis.
 Time intervals when the LAT boresight was rocked with respect to the local zenith by more than $52^{\circ}$ 
 (usually for calibration purposes or to point at specific sources) and events with a reconstructed angle with respect to the local zenith $>100^{\circ}$  were excluded. 
 The latter selection was necessary to limit the contamination from $\gamma$ rays produced by interactions of cosmic rays with the upper atmosphere of the Earth. 
 In addition, to correct the calculation of the exposure for the zenith cut, time intervals when any part of the ROI was observed at zenith angles $>100^{\circ}$ were excluded.

To derive the spectral fluxes we applied an unbinned maximum likelihood technique to events in the energy range from 100\,MeV to 300\,GeV \citep{mattox96}.
Sources from the 2FGL catalog \citep{nolan12} located within $15^{\circ}$ of \pks\
were included in the model of the ROI by setting the spectral shapes and the initial parameters for the modeling to those published in the 2FGL catalog. 
In the fitting procedure the parameters of sources located within a $7^{\circ}$ ROI, as well as the normalization of 
 the isotropic background and the Galactic diffuse emission components,
 were left free to vary.
Parameters of sources located between $7^{\circ}$ and  $15^{\circ}$ from the source of interest were instead fixed at their catalog values. 
Instrumental systematic errors are typically $\sim$10\% and negligible compared to the large flux variations observed \citep{LATperform}.

During the LAT observation period, \pks\ was not always significantly detected. Consequently, we calculated flux upper limits at the 95\% confidence level for each time bin where the Test Statistic\footnote{The Test Statistic value quantifies the probability of having a point $\gamma$-ray source at the location specified and corresponds roughly to the square of the standard deviation assuming one degree of freedom \citep{mattox96}.
It is defined as TS = $-2\log (L_0 /  L)$, 
where $L_0$ is the maximum likelihood value for a model without an additional source (the 'null hypothesis') and $L$ is the maximum likelihood value for a model with the additional source at the specified location.
}
(TS) value for the source was TS$<$10 or the number of predicted photons was $N_{\mathrm{pred}}<3$.  The LAT flux results are shown in the bottom panel of Fig. \ref{mwllc}.  
The binning used for the LAT light curves provides a good compromise between sufficient photon statistics and sensitivity to source flux variability. We explored smaller time bins for the flares, but the poor statistics did not allow sufficient detections, even during bright flares.

\subsection{\emph{Swift} observations}

The \emph{Swift} satellite \citep{Gehrels2004} performed 6 target of opportunity
(ToO) observations on \pks\ in  2010 May, triggered by high optical
activity of the source~\citep{ATOMATel10}, and 2 ToO observations in 2011 May, 
triggered by the third $\gamma$-ray flare observed by \fermilat~\citep{ATEL3329}. 
To investigate the source behaviour
over the years we also analysed three \emph{Swift} observations carried out before the
launch of \fermilat (2005 April and 2006 May) and another three observations
in 2010 February and September. 
The observations were made with all three 
on-board instruments: the UV/Optical Telescope (UVOT; \citealt{Roming2005}, 170--600 nm), the X-ray Telescope (XRT; \citealt{Burrows2005}, 0.2--10.0 keV), and the Burst Alert Telescope (BAT; \citealt{Barthelmy2005}, 15--150 keV). The hard X-ray flux of this source is below the sensitivity of the BAT instrument, not appearing in the 70-month BAT catalog \citep{baumgartner}.

The XRT data were processed with standard procedures (xrtpipeline v0.12.6), filtering, and screening criteria by using the Heasoft package (v.6.11). The
source count rate was low during all the observations (count rate $<$ 0.5 counts s$^{-1}$ in the 0.3--10 keV energy range), thus we only considered
photon counting data for our analysis, and further selected XRT event grades 0--12. Pile-up correction was not required. Source events were
extracted from a circular region with a radius of  20 or 25 pixels (1
pixel $\sim$ 2.36$\arcsec$), depending on the source count rate, while
background events were extracted from a circular region with radius 50 pixels and located 
away from background sources. Ancillary response files were generated with the
task {\tt xrtmkarf}. These account for different extraction regions, vignetting
and point spread function corrections. We used the latest  version of the spectral redistribution
matrices in the calibration database maintained by HEASARC. The adopted energy range
for spectral fitting is 0.3--10 keV. We summed two observations performed on
 2010 February 11 in order to achieve higher statistics. When the number of counts was fewer 
 than 200 the Cash statistic \citep{Cash1979} on ungrouped data was used. 
All the other spectra were rebinned with a minimum of 20 counts per energy bin to allow $\chi{^2}$ fitting within
{\sc XSPEC} (v12.6.0; \citealt{Arnaud1996}). We fit the individual spectra with a simple absorbed power law, with a
neutral hydrogen column density fixed to its Galactic value ($N_{\rm H} = 7.71 \times$ 10$^{20}$
cm$^{-2}$; \citealt{Kalberla2005}). The fit results are reported in Table \ref{1424_XRT}. 
\begin{table*}[th!]
\footnotesize
\caption{Fitting results of {\it Swift}/XRT observations of \pks. 
Columns report, from left to right: observation time, net exposure time, 
observed photon index and flux. 
The last column indicates the method used to perform the spectral analysis:
reduced $\chi^2$ and, in parentheses, the degrees of freedom, 
or the Cash method when the statistics were low.
Results were obtained considering an absorbed power-law model with  $N_{\rm H}$ fixed to Galactic
absorption in the direction of the source. $^{a}$ Observed flux.}
\centering
\begin{tabular}{llcccc}
\hline
\hline
\noalign{\smallskip}
\multicolumn{1}{c}{Time} &
\multicolumn{1}{l}{Time } &
\multicolumn{1}{c}{Net Exp. Time} &
\multicolumn{1}{c}{Photon Index} &
\multicolumn{1}{c}{Flux$^{a}$ 0.3$-$10.0 keV} &
\multicolumn{1}{c}{$ \chi^{2}_{\rm red}$ (d.o.f.) / Cash} \\
 \multicolumn{1}{c}{(UT)} &
 \multicolumn{1}{l}{(MJD)} &
 \multicolumn{1}{c}{ (sec) } &
\multicolumn{1}{c}{}&
\multicolumn{1}{c}{(10$^{-12}$ erg cm$^{-2}$ s$^{-1}$)} &
 \\
\multicolumn{1}{c}{} \\
\hline
\noalign{\smallskip}
2005-Apr-19& 53479 & 2249 & $1.35 \pm 0.22$ & $3.34 \pm 0.62$ & Cash \\
2005-Apr-23&  53483 & 1543 & $1.54 \pm 0.41$ & $1.73 \pm 0.47$ & Cash \\
2006-Jun-18& 53904  & 3784 & $1.37 \pm 0.17$ & $4.28 \pm 0.49$ & 0.727 (11) \\
2010-Feb-11& 55238 & 3646 & $1.49 \pm 0.24$ & $3.77 \pm 0.75$ & 0.762(9) \\
2010-May-12& 55328 & 1958 & $1.70 \pm 0.21$ & $5.80 \pm 0.90$ & 0.796 (9) \\
2010-May-13& 55329 & 1279 & $1.75 \pm 0.24$ & $4.25 \pm 0.84$ & Cash \\
2010-May-14& 55330.5 & 1963 & $1.99 \pm 0.20$ & $5.25 \pm 0.63$ & 0.803 (10) \\
2010-May-14& 55330.6 &  951 & $1.78 \pm 0.20$ & $7.89 \pm 0.72$ & Cash \\
2010-May-15& 55331 & 1938 & $1.68 \pm 0.21$ & $4.45 \pm 0.71$ & Cash \\
2010-May-16& 55332 & 3833 & $1.82 \pm 0.18$ & $4.29 \pm 0.54$ & 0.868 (13) \\
2010-Sep-30& 55469 & 5152 & $1.47 \pm 0.15$ & $3.80 \pm 0.41$ & 1.008 (15) \\
2011-May-07& 55688 & 3913 & $1.80 \pm 0.24$ & $3.41 \pm 0.55$ & 0.920 (10) \\
2011-May-10& 55691 & 3932 & $1.61 \pm 0.23$ & $3.93 \pm 0.46$ & 1.080 (9) \\

\noalign{\smallskip}
\hline

\noalign{\smallskip}
\end{tabular}
\\
\label{1424_XRT}
\end{table*}
The average X-ray flux observed in  2010 mid-May was higher
than the values observed by {\it Swift}/XRT in 2005-2006 and
 2010 February, indicating an increase in X-ray activity and implying that the
flaring mechanism also influences the X-ray band. In particular, an increase of a factor of $\sim$2.5  was observed on 2010 May 14 compared to the 2005-2006 average, at the peak of the X-ray emission, which coincides with flare B. We noted also a change of the spectral index from $\sim$1.4 to $\sim$1.8
during  2010 May. On the other hand, no increase of the X-ray flux was observed after the 2011 May
 $\gamma$-ray flare. 

All pointings of UVOT in 2005 and 2006 as well as in 2010 February and September  were performed with all 6 UVOT filters (V, B, U, UVW1, UVM2, and UVW2). The remaining observations in 2010 were taken using the ``filter of the day", i.e., either the U or one of the UV filters. In 2011, the source was observed with the V filter only.

We re-processed the UVOT data using the script uvotgrblc, available in version 6.10 of the HEASoft software, and  version 20100930 of the UVOT calibration database. The orbits of each UVOT image were summed in order to increase the signal to noise ratio. The photometry was obtained by customizing the background region, selecting an annulus with inner/outer radius of 27\arcsec/35\arcsec, respectively. All field sources that contaminate the background region and appear in any filter have been masked out.
The  2005 April pointings had relatively short exposures and the chosen source extraction region was a 3\arcsec-radius circle. Since the source intensity is higher than the values in the remaining observations due to the longer exposures, the radius has been increased to 5\arcsec. The script estimates the photometry by calling the task uvotsource, and the output values are then corrected for aperture effects.
Flux values were
de-reddened using the values of E(B-V) taken from \citet{schlegel98} with $\mathrm{A_\lambda}$/E(B-V) ratios calculated for the UVOT filters using the mean interstellar extinction curve from \citet{fitzpatrick}.  The {\it Swift} UVOT and XRT results are shown in panel 4 and 5 of Fig. \ref{mwllc}.

\subsection{Radio data}

\begin{figure}
\centering
\resizebox{6cm}{!}{\rotatebox[]{-90}{\includegraphics{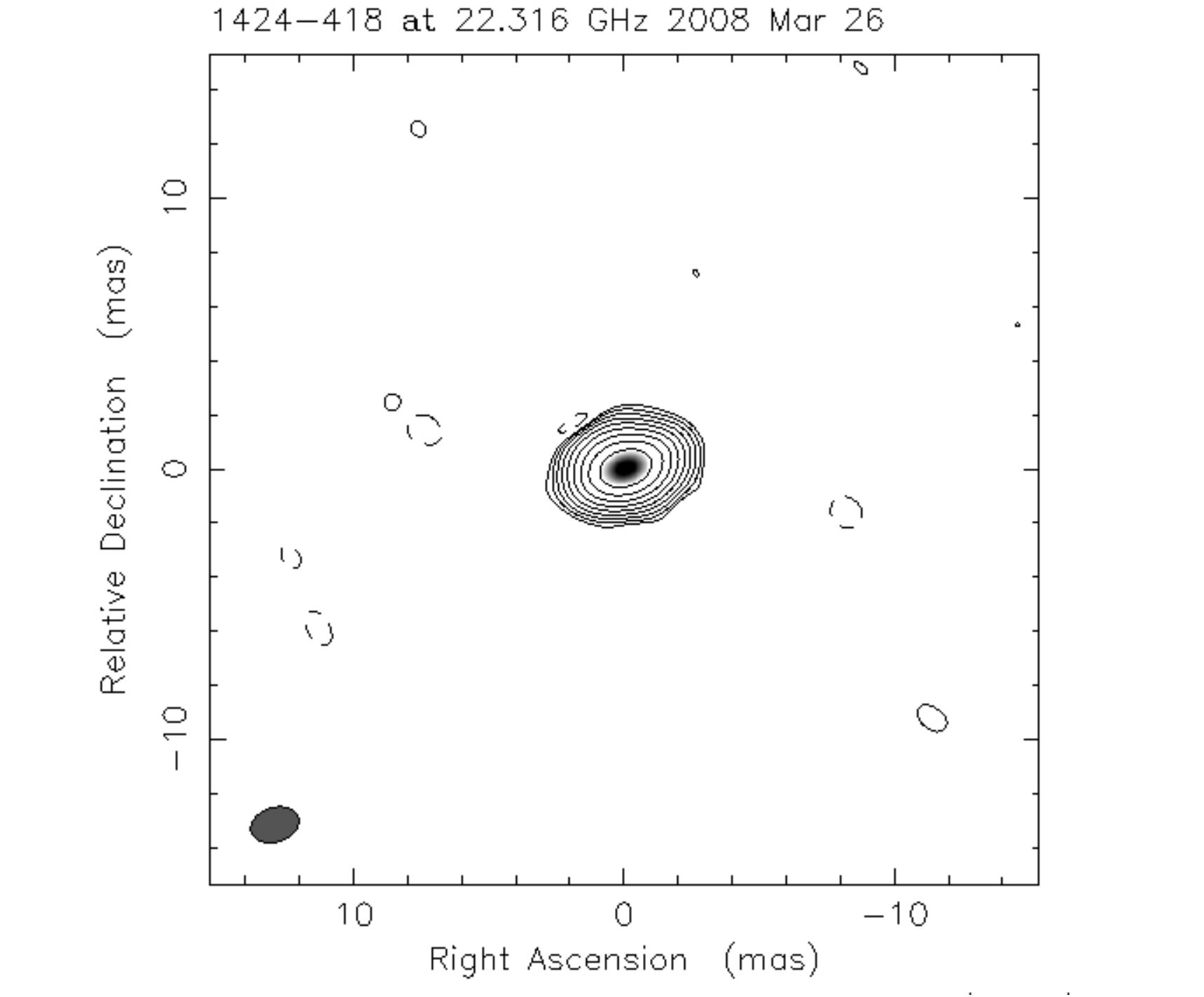}}}
\vspace{-13pt}
\caption{22.3\,GHz VLBI image of \pks\ confirming its core-dominated morphology. The image has a peak flux density of 0.8 Jy/beam. The hatched ellipse 
on the bottom left represents the synthesized beam of the observing array.}
\label{fig:1424-418_tanami26mar2008}
\end{figure}

High resolution Very Long Baseline Interferometry (VLBI) observations of this blazar carried out
as part of the TANAMI program show it to have a faint
low-surface-brightness jet with a wide opening angle \citep{Ojha2010}.
At the milliarcsecond scale, the 22.3\,GHz image of \pks\ (Fig.  \ref{fig:1424-418_tanami26mar2008}; 
observed on  2008 March 26) clearly indicates that this source is extremely core dominated at this
frequency. This is further confirmed by the  8.4\,GHz image in \citet{Ojha2010} 
which shows a dominant, compact VLBI core and a very diffuse and resolved jet.

\begin{table}
\caption{Flux densities of the milliarcsecond core of \pks.}
\centering
\begin{tabular}{cccc}
\hline \hline
Time & Time & Frequency  & Core Flux Density \\
 (UT) &  (MJD) & (GHz) & (Jy) \\
\hline 
2008-Aug-08 &  54686 & 8.4 & 1.5 \\
2009-Feb-23 & 54885 & 8.4 & 1.1\\
2010-Mar-12 & 55267 & 8.4 & 1.1\\
2010-Jul-24 &  55401 & 8.4 & 1.2\\
\hline
\end{tabular}
\label{tab:tanami}
\end{table}

The milliarcsecond core flux densities of \pks\ at
8.4\,GHz at four epochs during 2008 through 2010 are listed in Table  \ref{tab:tanami}. 
The core flux density 
declined after the  2008 August epoch and has remained steady during the
period covered by the flares. This general trend is seen in the lower
resolution radio data as well (see below). The errors in these flux densities 
are conservatively estimated to be less than 20\%. These flux densities were
obtained by model fitting a circular Gaussian to the core of VLBI
images made by the TANAMI program. For details on the observations,
imaging and model fitting process refer to \cite{Ojha2010}.

%
As part of the TANAMI program, the Australia Telescope Compact Array
(ATCA) was used
to make ``snapshot'' observations of \pks\ at frequencies
between 4.8 and 40\,GHz. Data at all frequencies were calibrated
against the ATCA primary flux calibrator PKS\,1934$-$638 \citep[see ][]{Stevens2012}.
These flux densities have a $1\sigma$  uncertainty of 5, 10 and 15\% at 4.8/9.0\,GHz, 19\,GHz, 
and 40\,GHz, respectively. Each frequency is the center of a 2\,GHz wide band.
%
%
%
\pks\ was also monitored at a
frequency of 6.7\,GHz by the 30-meter Ceduna radio telescope in South
Australia. Each flux density has a 1 $\sigma$ uncertainty of  $\pm
0.3$Jy \citep{McCulloch2005}. The lower-frequency radio results from the TANAMI program are shown in the top panel of Fig. \ref{mwllc}.

\subsection{SMA observations}
Observations at 230 GHz (1.3 mm) were obtained at the Submillimeter Array (SMA) 
near the summit of Mauna Kea (Hawaii).  
\pks\ is included in an ongoing monitoring program at the SMA to
determine the fluxes of compact extragalactic 
radio sources that can be used as calibrators at mm wavelengths \citep{Gurwell07}.   
\pks\ was also observed as part of a dedicated program to follow sources on the 
\fermilat Monitored Source List (PI: A. Wehrle).  

These potential calibrators are observed for 3 to 5 minutes, 
and the measured source signal strength is calibrated against known standards, 
typically solar system objects (Titan, Uranus, Neptune, or Callisto).  
Data from this program are updated regularly and are available at the SMA website\footnote{http://sma1.sma.hawaii.edu/callist/callist.html}.

\subsection{APEX observations} 
As part of the F-GAMMA program (e.g. \citealt{Fuhrmann2007}, \citealt{Angelakis2008}, \citealt{Angelakis2012}), 
sub-mm observations of a large sample of \fermilat $\gamma$-ray blazars including \pks\
have been performed with the APEX (The Atacama Pathfinder EXperiment) telescope in Chile since early 2008 (see also \citealt{Larsson2012}). The quasi-regular F-GAMMA
observations are obtained during several dedicated 
time-blocks per year complemented by regular and frequent pointing observations within the framework of other projects and APEX technical time.

The multi-channel bolometer array facility instrument LABOCA 
\citep[Large Apex BOlometer CAmera,][]{Siringo2008} used for these observations consists of 295 channels arranged in 9
concentric hexagons and operates at a wavelength of 0.87\,mm
(345\,GHz). The observations are typically performed in `spiral
observing mode' with a raster of four spiral maps each of 20 or 35
seconds integration, depending on the source brightness at
345\,GHz. During each run, Skydip\footnote{Skydips are measurements 
of the sky temperature as a function of airmass and are used 
 to estimate the zenith sky opacity (and so apply to astronomical data for the atmospheric extinction).}
measurements for opacity correction and
frequent calibrator measurements are performed 
(Fuhrmann et al. in prep.). 
The SMA and APEX flux results are shown in the second panel of Fig. \ref{mwllc}.




\section{Multiwavelength behaviour of the source}\label{sec:lc}


The most striking feature of the multiwavelength light curve of Fig. \ref{mwllc} is the series of strong, isolated flares present in the optical and $\gamma$-ray bands,  seen in the third and bottom panels respectively.
At these wavebands the source underwent two main flaring episodes,
the first occurring between 2009 June 22 to July 20 (MJD 55004 - 55032; flare A)
and the second between 2010 May 6 to 25 (MJD 55322 - 55341;  flare B).
A third, lower-amplitude flare
occurred between 2011 April 19 - May 16 (MJD 55670 - 55697; flare C).   In both bands, \pks\ remained in a relatively low state at other times.  A detailed analysis of these optical/$\gamma$-ray flares is presented in the following section. 

Lower-frequency radio flux densities are plotted in the
first panel of Fig. \ref{mwllc}.
Although the radio data indicate that the source has been almost steady
during the period of the flares, we note that the
Ceduna values (gray points) have quite large error bars, while the
sparse sampling of ATCA data may be consistent with some variability,
at least on long time scales.
%

At mm and sub-mm wavelength,  the overall sampling  displayed in the 
second panel of Fig. \ref{mwllc} (SMA and APEX data are represented by red and black symbols 
respectively) is limited, but a strong flux density increase is evident over the observing
period of more than two years: the overall
flux density increased by a factor of about 3, whereas faster, under-sampled 
variability is superimposed on the long-term increasing trend. 

\pks\ shows dramatic variability in the X-ray band and in all of the UVOT bands. The data are sparse 
and do not allow  firm conclusions about long term trends but they show that immediately 
after all three of the $\gamma$-ray flares, the optical/UV emission increased as well, 
when compared with periods of lower $\gamma$-ray activity. In particular, the source is $\sim 1-1.5$ and $\sim 0.5$ magnitude brighter in  2010 May and 2011 May, respectively, than when observed in 2010 February and September. 
During the latter periods, the intensity was moderately higher ($\sim 0.5$
magnitudes) than that observed in  2006 June, and significantly higher ($\sim 2$ magnitudes) than the very low level recorded in 2005 April. In X-rays, an increase of a factor of 2 in the \textit{Swift}-XRT flux was seen during 2010 May,
in coincidence with the second $\gamma$-ray flare.

 \section{Optical/gamma-ray correlation}\label{sec:var}
%
 \begin{figure*}[!ht]
\centering
\resizebox{13cm}{!}{\rotatebox[]{0}{\includegraphics[width=117px,height=70px]{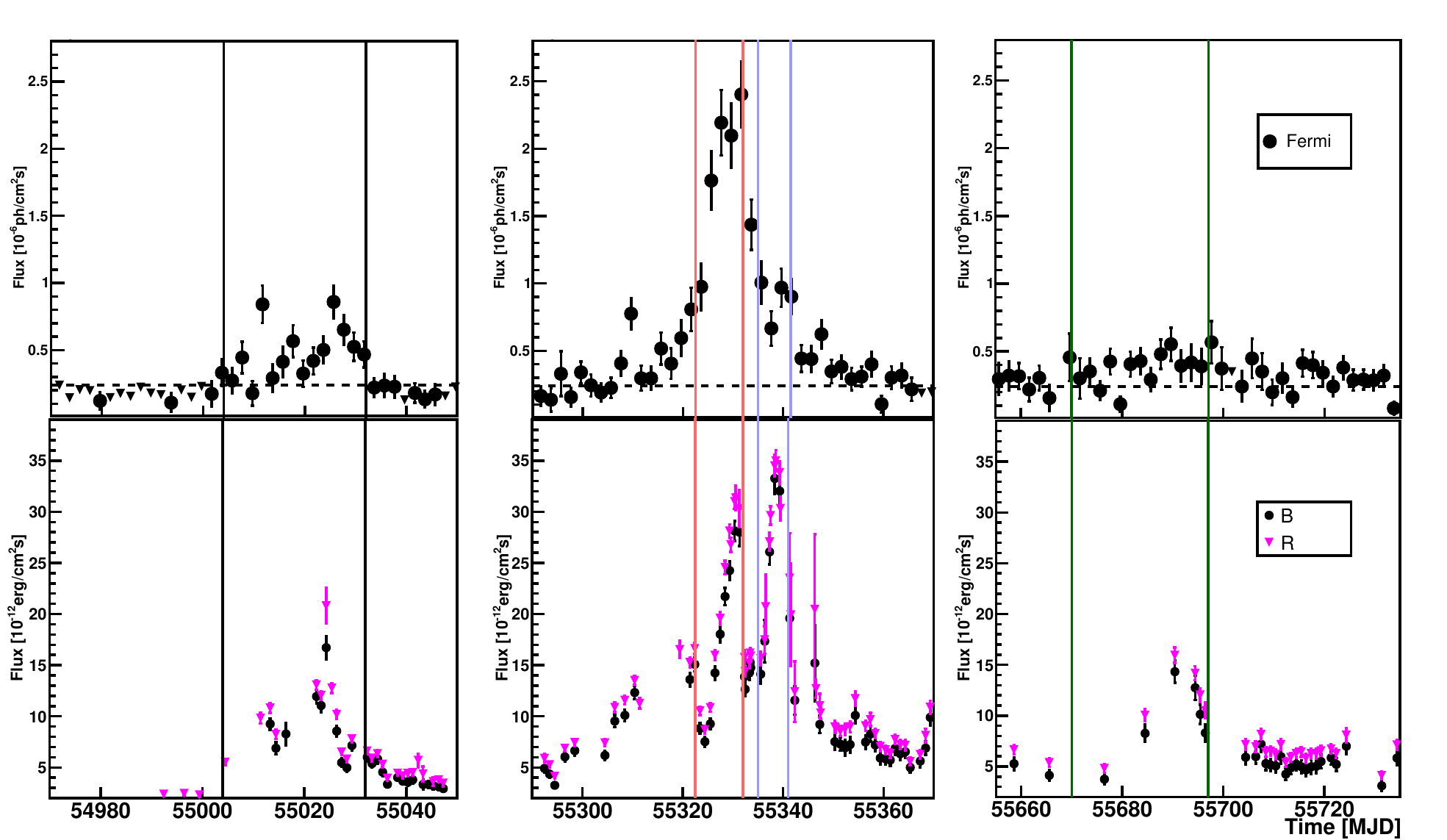}}}\\
 \caption{\fermilat (2-day bin, top panel) and optical light curves (bottom panel) of \pks.
 The dashed horizontal line in the upper panel indicates the mean $\gamma$-ray flux over the whole observing period considered in this work.
 Vertical lines denote the three prominent outbursts
 studied in this work:  black, flare A  (2009 June 22 - July 20, MJD 55004-55032), red, flare B1 (2010 May 6-16, MJD 55322-55332), blue, flare B2 (2010 May 19-25, MJD 55335-55341), and green, flare C (2011 April 19 - May 16, MJD 55670-55697).}
              \label{fermiAtomLc}%
 \end{figure*}
The three main outbursts are clearly visible in the 2-day bin $\gamma$-ray light curve and the optical light curves, Fig. \ref{fermiAtomLc}.
Flare A and flare B are displayed  with more detail 
in the upper panels of Fig. \ref{fig:Flare1LC} and \ref{fig:Flare2LC},
in which the $\gamma$-ray light curves (black crosses) have been superimposed onto the optical
flux values (pink points).

Close examination of the $\gamma$-ray flux evolution during flare B
shows a single flaring event. However, in 
 the optical band a two-peak sub-structure is visible during
the same time period. 
In our work, therefore, we study separately  the sub-periods
and refer to them as flare B1 (2010 May 6 -- 16, MJD 55322 -- 55332) and flare B2 (2010 May 19 -- 25; MJD 55335 -- 55341).

%
%

\subsection{Flare A}\label{subsec:a}
\begin{figure}[t]
\centering
\resizebox{9	cm}{!}{\rotatebox[]{0}{\includegraphics{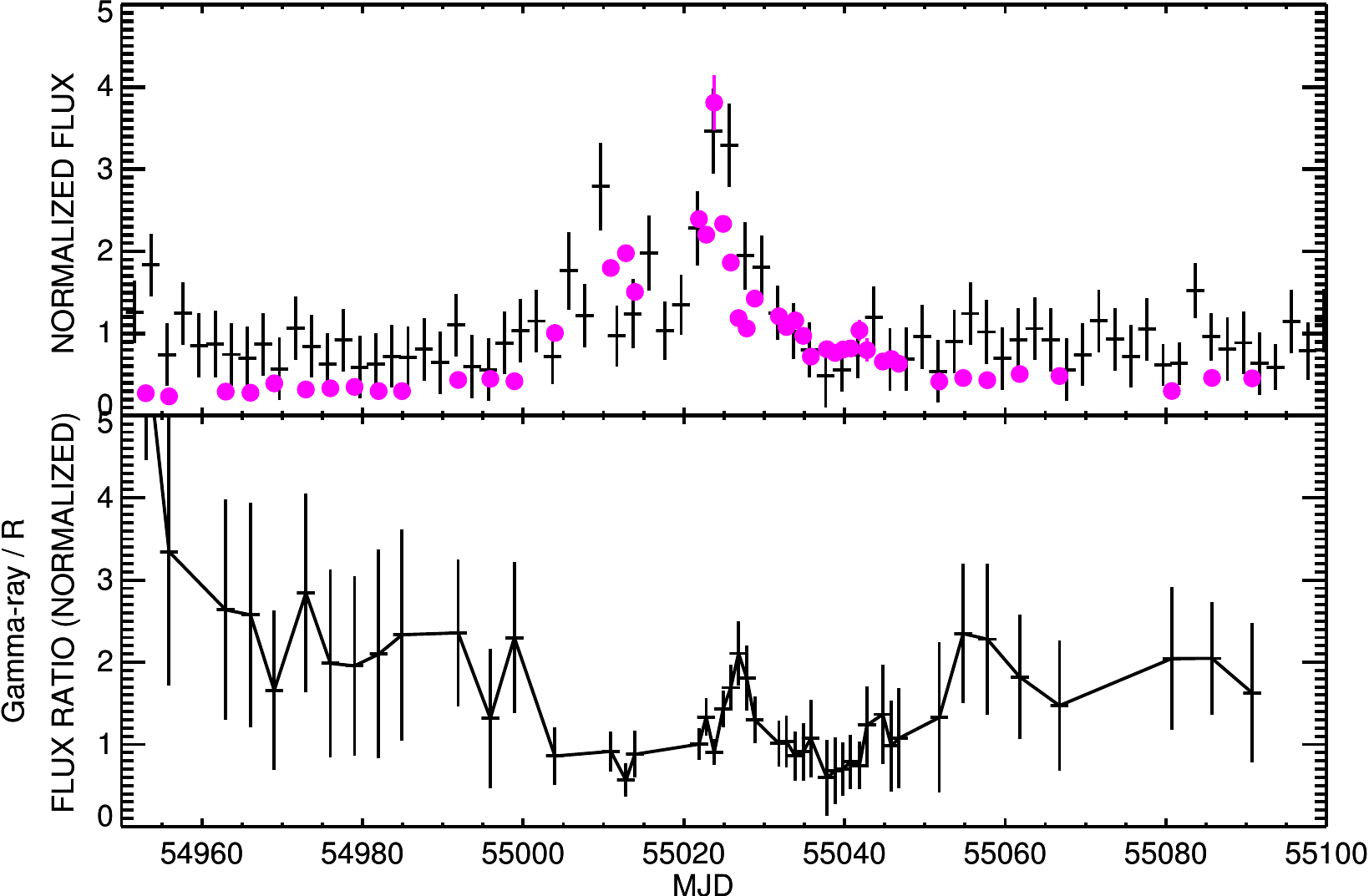}}}\\
\caption{\textit{Top panel}: 
Comparison of LAT (crosses) and R-band (filled circles) light curves for flare A. 
Fluxes are rescaled so the mean flux values  over the full time range are approximately 1.
\textit{Bottom panel}: Flux
ratio for the two arbitrarily normalized light curves computed by interpolating
LAT fluxes to the times of the R-band measurements.}
\label{fig:Flare1LC}
\end{figure}

The shape of flare A is very similar in the optical and $\gamma$-ray bands. As can be
seen in the upper panel of Fig. \ref{fig:Flare1LC}, not just the peak but also the rising and falling
branches seem to track each other fairly well. The main difference
between the two bands is a somewhat larger flare amplitude in
optical. This is apparent also in the flux ratio plot in
the lower panel of Fig. \ref{fig:Flare1LC}, although the comparison is
affected by the limited R band sampling during the first part
of the flare. The flux ratios 
were calculated by interpolating the LAT light curve to the
times of R-band measurements.  
To evaluate the ratio, given an optical flux value, we considered
two successive LAT flux values, before and after the optical one, 
and performed a linear interpolation between the two. 
This interpolated LAT flux is matched with the corresponding R-band measurement. 
Both light curves 
fluxes were rescaled so the mean flux values  over the full time range are approximately 1.
\begin{figure}[!ht]
\centering
\resizebox{9	cm}{!}{\rotatebox[]{0}{\includegraphics{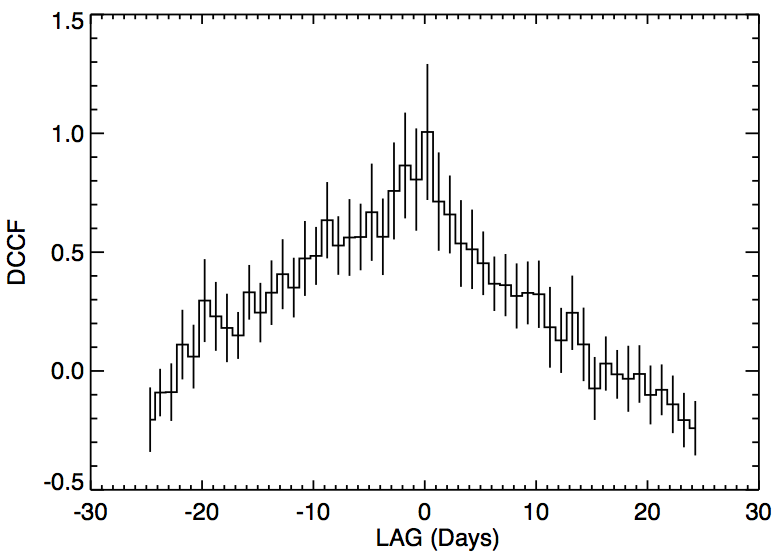}}}\\
\caption{Discrete Cross Correlation Functions for flare A using the ATOM R-band light curve (transformed to flux) and the 1-day bin LAT light curve ($>$ 100 MeV).}
\label{fig:DCCF1}
\end{figure}
The optical/$\gamma$-ray correlation 
was also investigated by calculating the Discrete Cross Correlation
Functions \citep[DCCF, ][]{edelson} between the R band flux and a 1-day binned LAT
light curve; 
 this is shown  in Fig. \ref{fig:DCCF1} for flare A.
Approximately 50 days of data were used for the correlation.
The DCCF is consistent with zero time lag. From a Gaussian fit we
 obtain a time lag = $-1 \pm 2$ days, where negative lag means optical leading
$\gamma$ rays. The uncertainty is based on Monte Carlo simulations taking
flux errors and time sampling into account \citep[see][]{Peterson1998}.
The rms variations in DCCF values for these Monte Carlo simulations
are shown as error bars in the plot, although we note that
these errors are strongly correlated:
they will make nearby bins in the DCCF move up or down together. 

\subsection{Flare B}\label{subsec:b}
\begin{figure}[b]
\centering
\resizebox{9	cm}{!}{\rotatebox[]{0}{\includegraphics{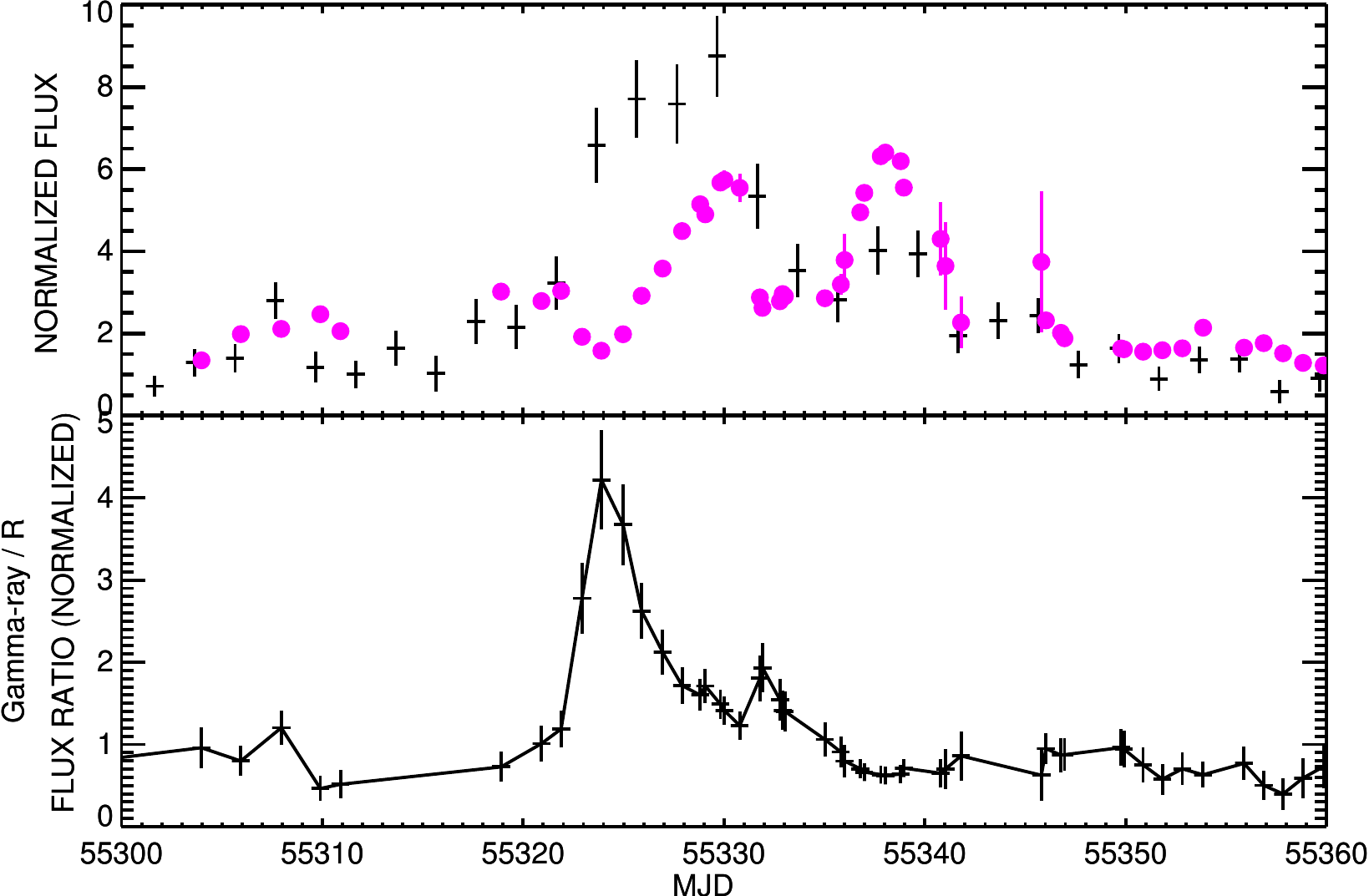}}}\\
\caption{\textit{Top panel}: 
Comparison of LAT (crosses) and R-band (filled circles) light curves for flare B, 
Fluxes are rescaled so the mean flux values  over the full time range are approximately 1.
\textit{Bottom panel}: Flux ratio computed by interpolating LAT fluxes to the times of the R-band 
measurements as calculated for flare A in Fig. \ref{fig:Flare1LC}.
Symbols and normalization are the same as those used for flare A in Fig. \ref{fig:Flare1LC}.}
\label{fig:Flare2LC}
\end{figure}
In contrast to the first flare, Flare B shows large differences between the two bands. 
Two separate flare
components are seen in optical but only one of these has a prominent
counterpart in the LAT light curve. The flare onset is also different
with a sharp, less than 1 day, increase in $\gamma$ rays and a much more
gradual brightening in optical. These differences are illustrated by
the relative flux of the two bands as plotted in the lower part of
Fig. \ref{fig:Flare2LC}, which refers to the period of Flare B and displays in the upper
panel the superposition of the optical and $\gamma$-ray light curves. 
The lower panel shows their flux ratios with the same normalization that was applied in Fig. \ref{fig:Flare1LC}.

\begin{figure}[!ht]
\centering
\resizebox{9	cm}{!}{\rotatebox[]{0}{\includegraphics{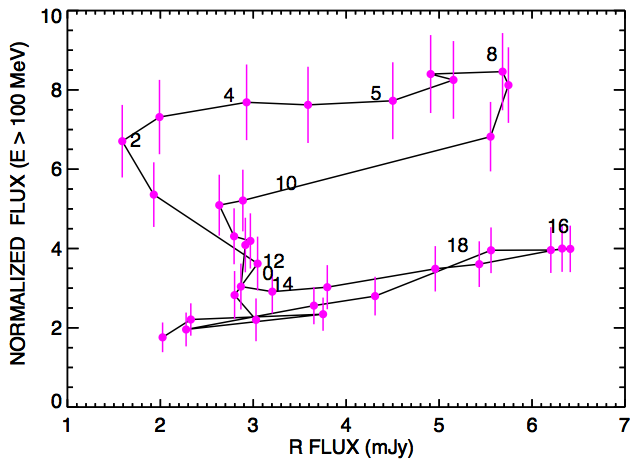}}}\\
\caption{
Gamma-ray/optical flux - flux evolution over flare B. LAT 2-day binned
fluxes are interpolated to the times of R-band measurements as explained 
in section \ref{subsec:a} and plotted on the y-axis. 
R-band fluxes are plotted on the x-axis. For clarity their errors 
(typically 5-10\% or less) are not shown.
Approximate times in days (from MJD 55322) are indicated along the track.
}	
\label{fig:Track}
\end{figure}
%
For the second flare, the relation between the two bands is more
complex than can be described by a single DCCF and cannot be explained by a single time lag.
We show in Fig. \ref{fig:Track} the flux-flux evolution for flare B.
The data points are the same values that were used to compute the
flux ratios in Fig. \ref{fig:Flare2LC}. In other words, $\gamma$-ray fluxes are
interpolated to the times of the R-band observations.
Approximate times in days (starting from MJD 55322) are indicated
along the track. The upper loop (days 0 - 12) in the plot corresponds to
the first of the two sub-flares (flare B1; 2010
May 6 -- 16, MJD 55322 -- 55332) and the lower part (from day 14) 
to the second sub-flare (flare B2; 2010 May 19 -- 25, MJD 55335 -- 55341), 
emitting predominantly in the R-band.
An interesting feature is that for the rise and decay of
the two optical sub-flares, the variation in $\gamma$-ray and
R-band fluxes can be described by a linear relation with similar
slope in all four cases (note, however, the data gap
in the R-band during the decay of the first sub-flare).
%

%
This behaviour can be interpreted as if there are two $\gamma$-ray components, 
one that is directly correlated with optical flux and one that is essentially unrelated
(the rise and final drop of the first ``sub-flare component'').
The former component contributes 
to both sub-flares, the latter is quiescent during sub-flare B2.
Further details are discussed with the SED modeling results in Section \ref{sec:sed}.

\section{Broad-band spectral energy distribution}\label{sec:sed}

\begin{figure*}[!ht]
\centering 
\includegraphics[width=.8\textwidth]{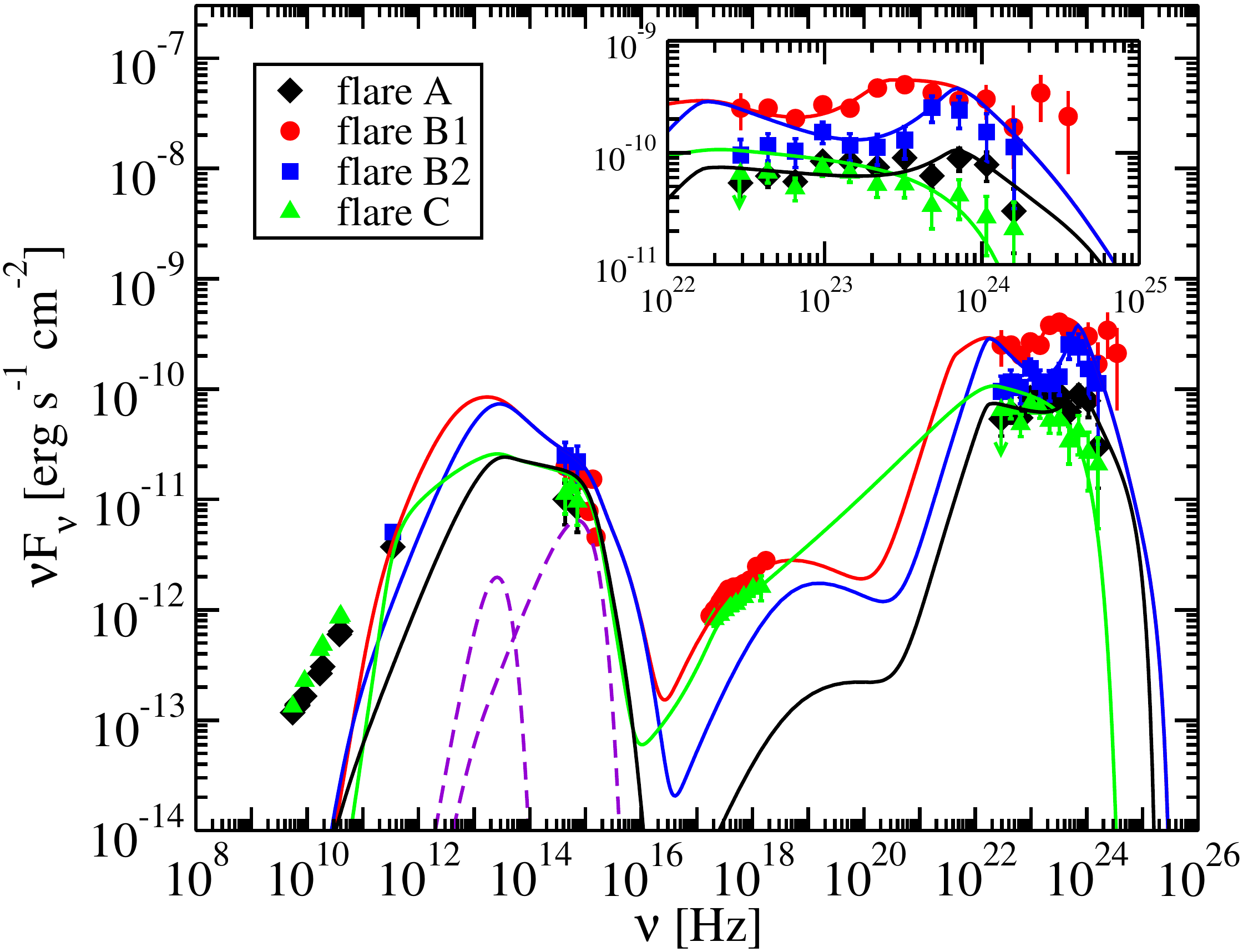}
\caption{SEDs and model fits for the four flares detected from \pks. 
Simultaneous data for flare A are represented by black symbols, for flare B1 by red symbols, 
for flare B2 by blue symbols and for flare C by green symbols. The dashed violet lines are the modeled spectra of the dust torus 
(peaked in the infrared) and accretion disk (peaked in the optical).  The solid lines are models of the total emission. 
The inset shows an enlargement of the LAT spectrum to point out the peculiar
concave upwards shape of  flares A, B1 and B2. 
} \label{SED_fig}
\end{figure*}

\begin{figure*}[t!]
\centering 
 \resizebox{12cm}{!}{\rotatebox[]{0}{\includegraphics{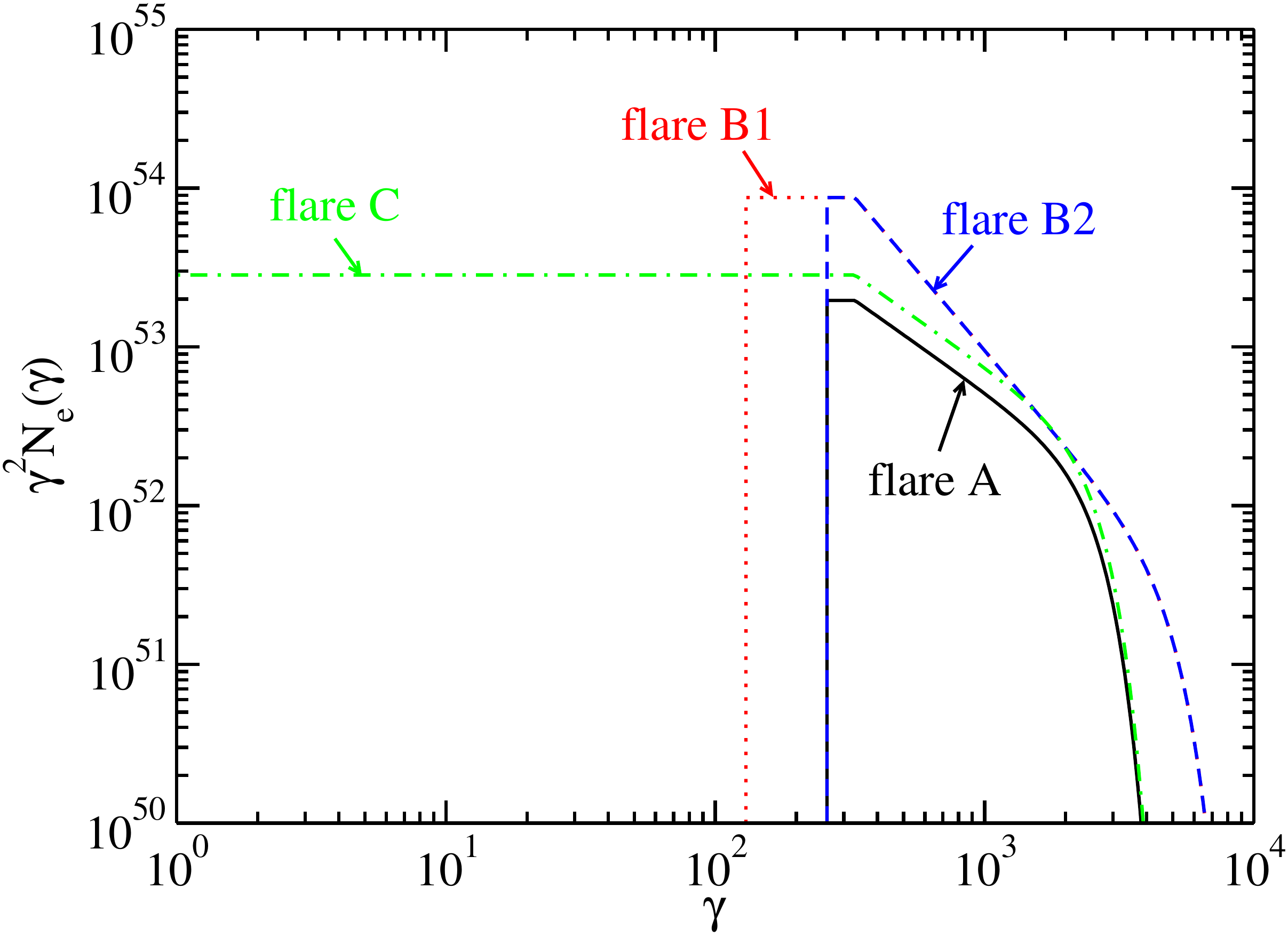}}}
\caption{Electron spectra (broken power laws) representing the four different flaring states considered for the source. Details on the parameter values are given in Table \ref{table_fit}.}
\label{fig:elecDistr}
\end{figure*}

We have built four SEDs around the four flare intervals using all multiwavelength data available.
The LAT spectra for flares A, B1 and B2  show a peculiar  upwards shape in the SED representation, although not statistically significant.
We  modeled each flare with
a leptonic model that includes the synchrotron, synchrotron
self-Compton (SSC), and external Compton (EC) processes.  Details of
the calculations can be found in \citet{finke08_SSC} and \citet{dermer09_EC}.  
As is common with blazar SEDs \citep[e.g. ][]{dammando12}, our model fit did not 
account for the bulk of the observed  radio flux densities. 
This emission, in the framework of the blazar scenario, must  be produced 
further down the jet, at relatively large distances from the blazar emission zone.
The SEDs are presented in Fig. \ref{SED_fig}
and the model parameters can be found in Table \ref{table_fit}.  The
electron distribution was assumed to be a broken power-law with a
super-exponential cutoff at high electron Lorentz factor $\gp$ 
(in the frame co-moving with the jet: $N_e \propto \g^{\prime -p_2}
\exp( -(\gp/\gp_{max})^4 )$ for $\gp > \gp_{brk}$).
 This electron distribution was chosen to fit the SEDs, and does not necessarily reflect particular acceleration or cooling processes.
%
Due to the odd concave upwards shape of the LAT spectra in the SED representation, the source was modeled with
two external seed photon sources, one with parameters similar to what
one would expect from a broad line region (BLR, \#1), and one similar to what one would
expect from a dust torus (\#2).  Both sources were modeled as
monochromatic sources that are isotropic in the host frame
(i.e., the frame of the host galaxy and black hole).  
In the host frame, the sources have energy densities $u_{seed,1}$ and $u_{seed,2}$ and dimensionless monochromatic energies $\epsilon_{seed,1}$ and $\epsilon_{seed,2}$.
The four different electron spectra derived from the modeling for the flares 
are shown in Fig. \ref{fig:elecDistr}.
The electron
spectra needed to be exceptionally narrow in order to explain the LAT
spectra.
In general, we attempted to fit as many of the flares as possible with the 
same parameters while only varying the electron distribution.
We succeeded in doing this for the flares A, B1, and
B2; the only difference in the model for these flares is the electron
distribution.  Flares B1 and B2 differ from flare A only by having a
harder spectrum above the break.  Flares B1 and B2 differ from each
other by flare B2 having a higher $\gp_{min}$ than flare B1.  
In this model, 
this explains why, although flares B1 and B2 have similar
optical behaviour and similar flux - flux evolution (see Fig. \ref{fig:Track}), 
flare B2 has lower $\gamma$-ray flux than flare B1, 
as seen in Fig. \ref{mwllc}.  
Flare C does not show the
concave upwards feature. This flare was modeled with only one external
seed photon source, the one representing the dust torus, and so it
represents a flare taking place outside of the BLR.  It also has a
lower $B$ and $\gp_{min}$ relative to the models for the other flares.
This was possible for flares A, B1 and B2 but not C. 

Note that \citet{finke10_3c454} suggested that the LAT spectra of
3C~454.3 can be modeled as a combination of EC from two seed photon
sources.  In that case, they used the accretion disk and BLR as their
photon sources.  The unusual shape of the LAT spectra in \pks\ 
indicates that a similar combination of seed photon sources
can be used to model this source for flares A, B1, and B2, although 
in this case we use the BLR and dust torus as the sources.  

We have also included a model for the accretion disk and dust torus
(dashed violet curves in Fig. \ref{SED_fig}). 
We used the black hole mass estimate of \citet{fan04} ($M_{BH} \approx 4.5\times10^9\
M_\odot$) in the accretion disk model, assumed to be a Shakura-Sunyaev disk \citep{shakura73}.
There is no evidence for a blue bump in the SED, so the disk was
modeled with enough luminosity that a fraction of this could explain
the EC seed photon sources.  If seed photon source \#1 represents the
BLR, and the BLR luminosity is related to the BLR radius by the
relation $L_{BLR}/(10^{45}\ \erg\ \s^{-1}) = [R_{BLR}/(10^{17}\
\cm)]^2$ \citep{ghisellini08}, then 
$L_{BLR}/L_{disk}
\approx 3\times10^{-3}$.  We also include an infrared bump representing the
dust torus, based on the parameters for seed photon source \#2.  In
this case we assumed a dust temperature $T_{dust}\approx 800$\ K and a
torus luminosity and radius that follow the relation
$L_{dust}/(10^{45}\ \erg\ \s^{-1}) = [R_{dust}/(2.5\times10^{18}\
\cm)]^2$.  With our models this implies 
$L_{dust}/L_{disk} \approx 0.4$.

We have computed the jet powers for the model fits as well
\citep[e.g.,][]{celotti93,finke08_SSC}, assuming a two-sided jet.
For flares A, B1, and B2, the result is quite far from equipartition.
The electrons have significantly less energy than the magnetic field,
due to the narrow electron spectra needed to fit the concave upwards
LAT spectra.  Flare C, on the other hand, has no such LAT spectrum,
and thus it is possible to fit it with a much broader electron
spectrum, and therefore is closer to equipartition.  A black hole mass
of $4.5\times10^9\ M_\odot$ implies an Eddington luminosity of
$L_{Edd} = 5.7\times10^{47}\ \erg\ \s^{-1}$, so $L_{disk} \approx 0.2
L_{Edd}$.  The jet powers of all of the models are about $P_{j,tot} =
P_{j,B}+P_{j,e}\approx 0.3 L_{Edd}$ so it appears approximately the
same amount of energy is going into the disk and jet.
%
%

%
%

\section{Summary and Conclusions}\label{sec:conclu}

We have presented multiwavelength observations of \pks\ during a period of 33 months (2008 August -- 2011 May) including  data from Ceduna, ATCA, SMA, APEX, ATOM, \textit{Swift} and \emph{Fermi} LAT.
Throughout the overall observing period significant variability is clearly present at optical and $\gamma$-ray frequencies, 
whereas only moderate variability can be noted at radio and sub-mm frequencies.
Focusing on the study of the optical and $\gamma$-ray behaviour of \pks , 
four main flaring phases have been pointed out and analyzed in detail. 
Good correlation is found between these energy bands during all periods with
the only exception of one flare (Flare B2).

%
The relative lack of variability in the VLBI, ATCA and Ceduna data
suggests that either the mechanism causing the optical-$\gamma$-ray flares is not
linked to the radio emission or that there is a delay before changes
in radio emission become evident. However, the relatively sparse radio monitoring data 
means that we cannot rule out radio variability on timescales shorter than the monitoring 
cadence as is seen in some other blazars \citep[e.g. ][]{richards}. The 
cadence of the Ceduna data is sufficient to rule out significant radio flares on $\sim 100$ day 
timescales.



We complemented the variability investigations building and modeling the spectral energy distribution  
 for flare A, flare B1, flare B2 and flare C.
These SEDs were fitted with a leptonic model which included the SSC and EC processes.
Based on the unusual LAT spectra for these flares, 
it appears that the $\gamma$-ray emission originates from 
the scattering of two external seed radiation fields: the dust torus and the BLR. 
In contrast to all other flare SEDs presented here, the SED of flare C is adequately modeled with only one EC component, 
the dust torus, and presents a slightly lower magnetic field value.
We find that in all outburst states the prevalent source of seed photons 
is consistent with a dust torus origin (with only about $4\%$ being provided by the BLR).
%
%
As noted, flare B shows a remarkably complex behaviour 
with a single evident flux increase at $\gamma$ rays
coincident in time with a double structured flare in the optical band.
Similar behaviour has already been reported for the blazars \object{4C\,+38.41} \citep{raiteri12} and 
\object{PKS 0208-512}. For the latter source  \citet{chatterjee13} suggested that changes in the magnetic field 
or in the bulk Lorentz factor could explain the absence of a $\gamma$-ray counterpart to the optical  outburst. 
However, the same speculations cannot be applied in the case of \pks.
Examination of the SEDs and the modeling results show that flares A, B1, and B2 can be explained by varying 
only the electron distribution. In particular, flares B1 and B2, which displayed different behaviour in the 
optical and $\gamma$-ray bands, have approximately the same optical brightening, but flare B1 is brighter in $\gamma$ rays.  
Looking at the $\gamma$-ray SEDs though, it can be seen that the LAT spectrum is at the same level for both flare B1 and flare B2, but that flare B2 has lower flux contribution around 
0.4 GeV to 2 GeV.  
Results of the SED modeling indicate that the value of $\gamma_{\mathrm{min}}$ is higher for flare B2 than flare B1:
this is the only difference between the two outbursts
and the only change needed to explain the difference between these two flaring states.

In conclusion, our investigation of multiple flares of \pks\ shows that, at least 
 in some objects, major variations in the overall blazar behaviour can be explained
 by changes in the flux and energy spectrum of the particles in the jet that are radiating.
 In this context, detailed studies of individual blazars like \pks\ 
 constitute an important opportunity to identify and unveil the fundamental mechanisms
 at work in blazar physics.

\begin{table*}[t]
\footnotesize
\caption{Model parameters for the SED shown in Fig.~\ref{SED_fig}
coincident with  the following flaring intervals: flare A (2009 June 22 - July 20, MJD 55004-55032), flare B1 (2010 May 6-16, MJD 55322-55332), flare B2 (2010 May 19-25, MJD 553325-55341) and flare C (2011 April 19 - May 16, MJD 55670-55697).
The first source of seed photons that has been taken into account is
 the BLR,  while the second seed source is the dust torus. 
 $^a$\,Seed photon energies are given in units of electron rest energy.} 
\label{table_fit}
\centering
\begin{tabular}{lccccc}
\hline \hline
Parameter & Symbol & Flare A & Flare B1 & Flare B2 & Flare C \\
 \hline
Redshift & 	$z$	& 1.522	& 1.522	& 1.522	& 1.522	  \\
Bulk Lorentz Factor & $\Gamma$	& 37	& 37 & 37 & 37	  \\
Doppler factor & $\delta_D$	& 37	& 37 & 37 & 37	  \\
Magnetic Field (G)& $B$         & 2.5 G  & 2.5 G & 2.5 G & 2.1 G   \\
Variability Timescale (s)& $t_v$       & 1.0$\times10^5$ & 1.0$\times10^5$ & 1.0$\times10^5$ & 1.0$\times10^5$  \\
Comoving radius of blob (cm)& $R^{\prime}_b$ & 4.4$\times$10$^{16}$ & 4.4$\times$10$^{16}$ & 4.4$\times$10$^{16}$ & 4.4$\times$10$^{16}$ \\
\hline
Low-Energy Electron Spectral Index & $p_1$       & 2.0 & 2.0 & 2.0 & 2.0     \\
High-Energy Electron Spectral Index  & $p_2$       & 3.2 & 4.0 & 4.0 & 3.2 	 \\
Minimum Electron Lorentz Factor & $\gamma^{\prime}_{min}$  & $2.6\times10^2$ & $1.3\times10^2$ & $2.6\times10^2$ & $1.0$   \\
Break Electron Lorentz Factor & $\gamma^{\prime}_{brk}$ & $3.3\times10^2$ & $3.3\times10^2$ & $3.3\times10^2$ & $3.3\times10^2$ \\
Maximum Electron Lorentz Factor & $\gamma^{\prime}_{max}$  & $2.6\times10^3$ & $5.0\times10^3$ & $5.0\times10^3$ & $2.6\times10^3$ \\
\hline
Black hole Mass ($M_\odot)$ & $M_{BH}$ & $4.5\times10^9$ & $4.5\times10^9$ & $4.5\times10^9$ & $4.5\times10^9$ \\
Disk luminosity ($\erg\ \s^{-1}$) & $L_{disk}$ & $1.0\times10^{47}$ & $1.0\times10^{47}$ & $1.0\times10^{47}$ & $1.0\times10^{47}$ \\
Inner disk radius ($R_g$) & $R_{in}$ & $6.0$ & $6.0$ & $6.0$ & $6.0$  \\
Seed ph. source \#1 energy density ($\erg\ \cm^{-3}$) & $u_{seed,1}$ & $2.2\times10^{-3}$ & $2.2\times10^{-3}$ & $2.2\times10^{-3}$ & 0.0 \\
Seed ph. source \#1 photon energy$^a$   & $\e_{seed,1}$ & $4.0\times10^{-5}$ & $4.0\times10^{-5}$ & $4.0\times10^{-5}$ & 0.0 \\
Seed ph. source \#2 energy density ($\erg\ \cm^{-3}$) & $u_{seed,2}$ & $5.5\times10^{-4}$ & $5.5\times10^{-4}$ & $5.5\times10^{-4}$ & $5.5\times10^{-4}$ \\
Seed ph. source \#2 photon energy$^a$ & $\e_{seed,2}$ & $4.0\times10^{-7}$ & $4.0\times10^{-7}$ & $4.0\times10^{-7}$ & $4.0\times10^{-7}$ \\
Dust Torus luminosity ($\erg\ \s^{-1}$) & $L_{dust}$ & $4.0\times10^{46}$ & $4.0\times10^{46}$ & $4.0\times10^{46}$ & $4.0\times10^{46}$   \\
Dust Torus radius (cm) & $R_{dust}$ & $4.1\times10^{18}$ & $4.1\times10^{18}$ & $4.1\times10^{18}$ & $4.1\times10^{18}$  \\
\hline
Jet Power in Magnetic Field ($\erg\ \s^{-1}$) & $P_{j,B}$ & $1.3\times10^{47}$ & $1.2\times10^{47}$ & $1.2\times10^{47}$ & $8.8\times10^{46}$ \\
Jet Power in Electrons ($\erg\ \s^{-1}$) & $P_{j,e}$ & $2.2\times10^{44}$ & $1.4\times10^{45}$ & $7.3\times10^{44}$ & $2.1\times10^{45}$ \\
\hline
\end{tabular}
\end{table*}

  \begin{acknowledgements}
The \fermilat Collaboration acknowledges generous ongoing support from a number 
of agencies and institutes that have supported both the development and the 
operation of the LAT as well as scientific data analysis.  These include the 
National Aeronautics and Space Administration and the Department of Energy 
in the United States, the Commissariat \`a l'Energie Atomique and the Centre 
National de la Recherche Scientifique / Institut National de Physique Nucl\'eaire 
et de Physique des Particules in France, the Agenzia Spaziale Italiana and the 
Istituto Nazionale di Fisica Nucleare in Italy, the Ministry of Education, 
Culture, Sports, Science and Technology (MEXT), High Energy Accelerator Research 
Organization (KEK) and Japan Aerospace Exploration Agency (JAXA) in Japan, and 
the K.~A.~Wallenberg Foundation, the Swedish Research Council and the Swedish 
National Space Board in Sweden.
Additional support for science analysis during the operations phase from the 
following agencies is also gratefully acknowledged: the Istituto Nazionale di 
Astrofisica in Italy and the K.~A.~Wallenberg Foundation in Sweden for providing 
a grant in support of a Royal Swedish Academy of Sciences Research fellowship for JC.
      Part of this work was supported by the German
      \emph{Deut\-sche For\-schungs\-ge\-mein\-schaft, DFG\/} project 
      number Ts~17/2--1.
The Australian Long Baseline Array and the Australia Telescope Compact
Array are part of the Australia Telescope National Facility which is
funded by the Commonwealth of Australia for operation as a National
Facility managed by CSIRO.
The Submillimeter Array is a joint project between the Smithsonian Astrophysical Observatory and the Academia Sinica Institute of Astronomy and Astrophysics and is funded by the Smithsonian Institution and the Academia Sinica.
This research was funded in part by NASA through {\it Fermi} Guest
Investigator grants NNH09ZDA001N and NNH10ZDA001N. This research was
supported by an appointment to the NASA Postdoctoral Program at the
Goddard Space Flight Center, administered by Oak Ridge Associated
Universities through a contract with NASA.
We thank Neil Gehrels and the \emph{Swift} team for scheduling our Target of Opportunity
requests. 
This research was enabled in part  through \emph{Swift} Guest Investigator
grants 6090777. 
We thank Silvia Rain\`{o}  for useful comments and suggestions.
\end{acknowledgements}

\bibliographystyle{aa} 
\bibliography{PKS1424m418} 
\end{document}